\documentclass[3p,times,twocolumn]{elsarticle}
 \biboptions{comma,sort&compress}
 
\usepackage{graphicx}
\usepackage{amsmath}
\usepackage{here}
\usepackage{stackrel}
\usepackage{amssymb,accents,latexsym,mathrsfs,bm,bbm}
\usepackage{ecrc}

\newcommand{\Tr}{{\textrm {Tr}}}

\newcommand{\U}{{\textrm {U}}}

\newcommand{\cA}{{\cal A}}  
\newcommand{\cB}{{\cal B}}

\newcommand{\cL}{{\cal L}}  
  
\newcommand{\cV}{{\cal V}}
\newcommand{\cW}{{\cal W}} 
 
\newcommand{\mcA}{{\mathcal A}}
\newcommand{\mcV}{{\mathcal V}}
\newcommand{\mbbV}{{\mathbb{V}}}
\newcommand{\mbhV}{{\mathscr{V}}}

\newcommand{\cons}{{\textrm{\scriptsize cons}}}
\newcommand{\cov}{{\textrm{\scriptsize cov}}}
\newcommand{\BZ}{{\textrm{\scriptsize BZ}}}

\newcommand{\eem}{{\textrm{\scriptsize em}}}

\volume{00}

\firstpage{1}

\journalname{Nuclear and Particle Physics Proceedings}
\runauth{Eugenio Meg\'{\i}as}

\jid{nppp}
\jnltitlelogo{Nuclear and Particle Physics Proceedings}

\usepackage{amssymb}

\usepackage[figuresright]{rotating}

\begin{document}

\begin{frontmatter}

\title{
%
Anomalous transport from equilibrium partition functions\,$^*$} 
 
 \cortext[cor0]{Talk presented at QCD22, 25th International Conference in QCD (04-07/07/2022, Montpellier - FR). }

 \author{Eugenio Meg\'{\i}as}
   \address{Departamento de F\'{\i}sica At\'omica, Molecular y Nuclear and Instituto Carlos I de F\'{\i}sica Te\'orica y Computacional, Universidad de Granada, Avenida de Fuente Nueva s/n, 18071 Granada, Spain}
\ead{emegias@ugr.es}

\pagestyle{myheadings}
\markright{ }
\begin{abstract}
\noindent
We summarize recent advances in the application of the equilibrium partition function formalism for the study of the transport coefficients of relativistic fluids induced by quantum anomalies, at first and second order in the hydrodynamic expansion. We provide results for theories with Abelian and non-Abelian chiral fermions, and discuss some features of the corresponding constitutive relations.
 
\begin{keyword}  Relativistic fluids, Hydrodynamics, Quantum anomalies, Anomalous transport.


\end{keyword}
\end{abstract}
\end{frontmatter}
\section{Introduction}

One of the most fruitful techniques to study out-of-equilibrium systems is the hydrodynamical approach, in which it is assumed local thermodynamical equilibrium. The hydrodynamical systems obey conservation laws of the energy-momentum tensor and charged currents, and the expectation values of these quantities are written in terms of fluid variables in the so-called constitutive relations, which are organized in a derivative expansion. In the presence of quantum anomalies the currents are no longer conserved, and this has important effects in the hydrodynamic description. In addition to the perfect fluid and dissipative contributions, new extra terms appear in the constitutive relations which turn out to be of non-dissipative nature, i.e. for the charged currents $\langle J^\mu \rangle = n u^\mu + \langle J^\mu \rangle_{\textrm{\scriptsize diss \& anom}}$. Two relevant phenomena appear at first order in the hydrodynamic derivative expansion as a consequence of chiral anomalies: the {\it chiral magnetic}~\cite{Fukushima:2008xe} and {\it chiral vortical}~\cite{Son:2009tf} {\it effects}. They consist in the generation of electric currents driven by and parallel to a magnetic field and a vorticity vector, respectively, i.e. $\langle J^\mu\rangle_{\textrm{\scriptsize anom}} = \sigma^\cB {\mathcal B}^\mu + \sigma^\cV \omega^\mu$. The corresponding susceptibilities are parity $({\cal P})$ odd and time reversal $(\cal T)$ even, the latter implying that they cannot contribute to entropy production, i.e. $\partial_t s_{\textrm{\scriptsize anom}} = 0$. These coefficients have been computed in a wide variety of methods, including kinetic theory~\cite{Stephanov:2012ki}, Kubo formulae~\cite{Landsteiner:2012kd}, fluid/gravity correspondence~\cite{Erdmenger:2008rm} and equilibrium partition function (EPF) formalism~\cite{Banerjee:2012iz}. In this work we will focus on the latter and address the study of the anomaly-induced contributions to the constitutive relations in both Abelian and non-Abelian gauge theories.

\section{Equilibrium partition function formalism}
\label{sec:Eq_part_func}

We begin by giving a brief summary of the EPF formalism introduced in~\cite{Banerjee:2012iz,Jensen:2013kka,Bhattacharyya:2013lha,Megias:2014mba}. Let us consider a relativistic invariant quantum field theory with a time independent $\U(1)$ gauge connection on the manifold
\begin{align}
ds^2 &= G_{\mu\nu} dx^\mu dx^\nu  \nonumber \\
     &=- e^{2\sigma(\vec{x})}(dt + a_i(\vec{x}) dx^i)^2 + g_{ij}(\vec{x}) dx^i dx^j \,, \\
{\cal A} &= {\cal A}_0(\vec{x}) dx^0 + {\cal A}_i(\vec{x}) dx^i \,.
\end{align}
The partition function of the system is defined as $Z = \Tr \, \exp\left(-\frac{H-\mu_0 Q}{T_0} \right)$, where $H$ is the Hamiltonian of the theory, and $Q$ is the conserved charged associated to the gauge connection, while $T_0$ and $\mu_0$ are the temperature and chemical potential at equilibrium. The dependence of the partition function on the fields, i.e. $\log Z = {\mathcal W}(e^\sigma, {\cal A}_0, a_i, {\cal A}_i, g^{ij},T_0,\mu_0)$, should be consistent with invariance under: i) 3-dim diffeomorphisms; ii) Kaluza-Klein (KK) transformation $[t \to t + \phi(\vec{x}) \,, \; a_i \to a_i - \partial_i \phi(\vec{x})]$; and iii) $\U(1)$ time-independent gauge transformations, up to gauge anomalies. In particular, KK invariance implies that the dependence in the gauge fields is only through the KK invariant combinations $A_0 \equiv \cA_0$ and $A_i \equiv \cA_i - a_i \cA_0$. From the partition function of the system, one can compute the energy-momentum tensor and consistent charged currents by performing appropriate $t$-independent variations. In doing that, one gets~\cite{Banerjee:2012iz}
\begin{eqnarray}
\hspace{-1.3cm}&&\langle J^i \rangle_\cons \!=\! \frac{T_0 e^{-\sigma}}{\sqrt{g_3}}\frac{\delta \mathcal{W}}{\delta A_i},  \qquad\quad\hspace{0.0cm} \langle J_0 \rangle_\cons \!=\! -\frac{T_0 e^{\sigma}}{\sqrt{g_3}}\frac{\delta \mathcal{W}}{\delta A_0}, \label{eq:cr_J} \\
\hspace{-1.3cm}&&\langle T_0^{\;i} \rangle \!=\! \frac{T_0 e^{-\sigma}}{\sqrt{g_3}}\left(\frac{\delta \cW}{\delta a_i} \!-\! A_0 \frac{\delta \cW}{\delta A_i}\right),  \langle T_{00} \rangle \!=\! -\frac{T_0 e^{\sigma}}{\sqrt{g_3}}\frac{\delta \cW}{\delta \sigma}, \label{eq:cr_T}
\end{eqnarray}
where $g_3 \equiv \det(g_{ij})$, and thus ${\mathcal W}$ plays the role of a generating functional for the hydrodynamic constitutive relations.

\section{Abelian anomalies and hydrodynamics}
\label{sec:Abelian}

We will present in this section the explicit results for the constitutive relations of a gas of massless Dirac fermions with $\U(1)$ gauge symmetry. The Lagrangian is
\begin{equation}
\qquad \cL = -i\overline\Psi\underline\gamma^\mu\nabla_\mu\Psi \,,
\end{equation}
where $\Psi = \left( \psi_L \; \psi_R \right)^T$ is a Dirac spinor, and $\nabla_\mu$ is the covariant derivative including the gauge field~$\cA_\mu$. The space-time dependent Dirac matrices are related to the Minkowski matrices by $\underline\gamma^\mu(x) = e^\mu_a(x) \gamma^a$, where $e^\mu_a(x)$ is the vierbein. We will study the properties of the EPF of this theory at first and second order in derivatives.

\subsection{Anomalous transport at first order}
\label{subsec:first_order}

The most general expression of the EPF at first order in the hydrodynamical expansion compatible with the symmetries mentioned in Sec.~\ref{sec:Eq_part_func}  is~\cite{Banerjee:2012iz,Megias:2014mba}
\begin{equation}
\hspace{-0.7cm}\cW_{(1)}  =  \!\!\int \!d^3x \!\sqrt{g_3}  \epsilon^{ijk} \Big[ \alpha_1 A_i F_{jk} + \alpha_2 A_i f_{jk} + \alpha_3 a_i f_{jk}  \Big]   \,, \label{eq:cW1}
\end{equation}
where $F_{ij} = \partial_i A_j - \partial_j A_i$ and $f_{ij} = \partial_i a_j - \partial_j a_i$, with coefficients $\alpha_i = \alpha_i(T,\nu)$ where $\nu \equiv \mu/T$, with $T = e^{-\sigma} T_0$ and $\mu = e^{-\sigma} A_0$ the out-of-equilibrium temperature and chemical potential, respectively. The $\U(1)$ current and energy-momentum tensor of the ideal gas of Dirac fermions write
\begin{eqnarray}
\hspace{-1cm}&&J^\mu = -\overline\Psi\underline\gamma^\mu  \Psi  \,,  \\
\hspace{-1cm}&&T_{\mu \nu} =  \frac{i}{4} \overline{\Psi} \left[ 
\underline{\gamma}_\mu \overrightarrow{\nabla}_\nu  - \overleftarrow{\nabla}_\nu \underline\gamma_\mu + (\mu \leftrightarrow \nu) \right ] \Psi \,.
\end{eqnarray}
The expectation values of $J^\mu$ and $T_{\mu\nu}$ at equilibrium may be computed from the thermal Green's function $\langle T\psi(-i\tau,  \vec{x})\psi^\dagger(0, \vec{x}^\prime) \rangle_\beta= T_0\sum_n e^{-i\omega_n\tau} \mathcal{G}( \vec{x},  \vec{x}^\prime,\omega_n)$, where $\omega_n =  2\pi T_0 \left( n + 1/2\right)$, and $T$ denotes time ordering. After performing the summation over Matsubara frequencies, one gets from a computation of $\langle J^i \rangle$ and $\langle T_0^{\;i} \rangle$ the following results for the chiral magnetic and chiral vortical conductivities
\begin{equation}
\sigma^\cB = C \mu \,, \qquad  \sigma^{\cV} = \frac{1}{2} C \mu^2 + C_2 T^2 \,,
\end{equation}
where the coefficients $C = 1/(4\pi^2)$ and  $C_2 = 1/24$ are induced by the chiral anomaly~\cite{Son:2009tf,Erdmenger:2008rm} and mixed gauge-gravitational anomaly~\cite{Landsteiner:2011cp,Landsteiner:2011iq}, respectively. These results have been obtained in a wide variety of methods, see e.g.~\cite{Fukushima:2008xe,Son:2009tf,Erdmenger:2008rm,Banerjee:2008th,Megias:2014mba,Landsteiner:2011cp,Landsteiner:2011iq}. Finally, by using the variational formulae (\ref{eq:cr_J})-(\ref{eq:cr_T}) with Eq.~(\ref{eq:cW1}), and after a comparison with the explicit expressions of the constitutive relations, one gets $\alpha_1(T,\nu) =  -\frac{C}{6} \nu$, $\alpha_2(T,\nu) =  -\frac{1}{2}\left( \frac{C}{6} \nu^2 - C_2 \right)$ and $\alpha_3(T,\nu) = 0$~\cite{Megias:2014mba}. For completeness, we present below the results for the constitutive relations in the theory with symmetry group $\U(1)_V \times \U(1)_A$, i.e. one vector and one axial current with chemical potentials $(\mu,\mu_5)$. These are given by~\cite{Landsteiner:2012kd}
\begin{eqnarray}
\hspace{-1.2cm}&&\langle J_a^\mu \rangle_{(1)} = (\sigma^\cB)_{a} \, {\mathcal B}^\mu + (\sigma^\cV)_a \, \omega^\mu \,, \quad (a = V, A) \,, \\
\hspace{-1.2cm}&&\langle T^{\mu\nu} \rangle_{(1)} =  u^\mu q^\nu + u^\nu q^\mu  \,, \quad q^\mu = \sigma^{\cB}_\varepsilon {\mathcal B}^\mu + \sigma^{\cV}_\varepsilon \omega^\mu \,, 
\end{eqnarray}
with
\begin{eqnarray}
\hspace{-1cm}&&(\sigma^\cB)_{V} = \frac{\mu_5}{2\pi^2} \,, \quad (\sigma^\cB)_{A} = \frac{\mu}{2\pi^2} \,, \\
\hspace{-1cm}&&(\sigma^\cV)_{V} = \frac{\mu \mu_5}{2\pi^2} \,, \quad (\sigma^\cV)_{A} = \frac{\mu^2 + \mu_5^2}{4\pi^2} + \frac{T^2}{12} \,, \\
\hspace{-1cm}&&\sigma^\cB_\varepsilon = (\sigma^\cV)_{V} \,, \quad \sigma^\cV_\varepsilon = \frac{\mu_5}{6\pi^2}(3\mu^2 + \mu_5^2) + \frac{\mu_5}{6} T^2 \,.
\end{eqnarray}
Here $(\sigma^\cB)_{V}$ is the chiral magnetic conductivity, $(\sigma^\cB)_{A}$ describes the generation of an axial current due to a magnetic field, and $(\sigma^\cV)_{V(A)}$ is the vector(axial) vortical conductivity. $\sigma^\cB_\varepsilon$ and $\sigma^\cV_\varepsilon$ are chiral magnetic and vortical conductivities for energy flux, respectively.

\subsection{Anomalous transport at second order}
\label{subsec:second_order}

Let us study the EPF at second order in the derivative expansion. The most general expression writes~\cite{Banerjee:2012iz}
\begin{eqnarray}
\hspace{-1.3cm}&&\cW_{(2)} = \int \!\! d^3 x \sqrt{g_3} \Bigl[ M_1 g^{i j} \partial_i T \partial_j T + 
  M_2 g^{i j} \partial_i \nu \partial_j \nu  \nonumber \\
\hspace{-1.3cm}&&+ M_3 g^{i j} \partial_i \nu \partial_j T +T_0^2 M_4 f_{i j} f^{i j} +    M_5 F_{i j} F^{i j} + T_0 M_6 f_{i j} F^{i j} \nonumber \\
\hspace{-1.3cm}&&+ M_7 \tilde R + N_1 \epsilon^{ijk} \partial_i A_0 f_{jk} + T_0^{-1} N_2 \epsilon^{ijk} \partial_i A_0 F_{jk} \Bigr] \,, \label{eq:W2}
\end{eqnarray}
where $\tilde R$ is the Ricci scalar in 3 dim, with $M_i = M_i(T,\nu)$ and $N_i = N_i(T,\nu)$. To get $\cW_{(2)}$ it is enough to compute $\langle J_0\rangle_{(2)}$ and $\langle T_{00}\rangle_{(2)}$ including only bilinear terms $\sim \partial_i X \partial_j Y$. The explicit expression of $M_7$ turns out to be $M_7 = -\frac{1}{144} T - \frac{1}{48\pi^2} T \,  \nu^2  +  \frac{1}{48 \pi^2} \frac{1}{T} M^2 \log 2$, where $M$ is the renormalization scale $(\bar M = 2^{-3/2} e^{\gamma_E} M)$. This coefficient is the relevant one for the computation of the transport coefficients presented below. The results for the rest of the coefficients in Eq.~(\ref{eq:W2}) are in Ref.~\cite{Megias:2014mba}. The terms proportional to $M^2$ can be renormalized by adding an appropriate counterterm. The renormalized effective action turns out to be not invariant under a Weyl rescaling due to the existence of terms $\propto \log \frac{\bar{M}^2}{T^2}$. Collecting these terms, one can identify the anomalous contribution to the partition function, a result that leads to the trace anomaly $\langle T_\mu^\mu \rangle =  -\frac{1}{24 \pi^2}  \mathcal{F}_{\mu \nu} \mathcal{F}^{\mu \nu}$~\cite{Giannotti:2008cv}.

The general result of the constitutive relations contains the following terms~\cite{Bhattacharyya:2013ida}
\begin{eqnarray}
\hspace{-1.35cm}&&\langle J_{\mu} \rangle_{(2)} \supset \upsilon_1 P_{\mu \alpha}  u_\nu R^{\nu \alpha}  + \upsilon_2 P_{\mu \alpha} \nabla_\nu \mathcal{F}^{\nu \alpha}  \,, \\ 
\hspace{-1.35cm}&&\langle T_{\mu \nu} \rangle_{(2)} \supset T \left(\kappa_1 R_{\langle \mu \nu \rangle} + \kappa_2 u^{\alpha} u^{\beta} R_{\langle \mu \alpha \nu\rangle \beta}  + \kappa_3 \nabla_{\langle \mu} \nabla_{\nu \rangle} \nu \right).
\end{eqnarray}
After using the variational formulae with $\cW_{(2)}$, one gets
\begin{equation}
\hspace{-0.5cm}\kappa_1 = \frac{T}{72} + \frac{1}{24\pi^2}\frac{\mu^2}{T} \,, \;\;\; \kappa_2 = 2 \kappa_1  \,, \;\;\; \kappa_3 =  -\frac{\mu}{12\pi^2}  \,.
\end{equation}
The results for $\upsilon_i$ are provided in Ref.~\cite{Megias:2014mba}. The results presented here for $\kappa_{1,2}$ are in agreement with Ref.~\cite{Moore:2012tc} after setting $\mu=0$. On the other hand, $\kappa_3$ and $\upsilon_2$ have been computed in a holographic model in 5 dim in Refs.~\cite{Erdmenger:2008rm,Banerjee:2008th,Megias:2013joa}, leading to the same parametric dependence for $\mu \ll T$. Finally, let us mention that the non-dissipative coefficients calculated above are ${\cal P}$-even and ${\cal T}$-even, while the second order coefficients that are ${\cal P}$-odd and ${\cal T}$-even vanish, i.e. $N_{1,2} = 0$.

\section{Non-Abelian anomalies and hydrodynamics}
\label{sec:Non_Abelian}

We will study in this section the constitutive relations within a theory with a non-Abelian chiral anomaly.

\subsection{The chiral anomaly}
\label{subsec:chiral_anomaly}

Let us consider a theory of chiral fermions with symmetry group $\U(N_f) \times \U(N_f)$, with Lagrangian
\begin{equation}
\hspace{-0.5cm} {\mathcal L} = i \overline\psi_L \gamma^\mu (\partial_\mu - i t_a \cA_{L\,\mu}^a) \psi_L +  i \overline\psi_R \gamma^\mu (\partial_\mu - i t_a \cA_{R\,\mu}^a) \psi_R  \,,
\end{equation}
where $t_a = t_a^\dagger$ are the Lie algebra generators. The chiral anomaly is signaled by the non-invariance of the effective action $i\Gamma = \cW$ under axial gauge transformations. This leads to the anomaly equation $\mathscr{A}_a(x) \Gamma[\mcV, \mcA] = G_a[\mcV, \mcA]$, where $G_a$ is the consistent anomaly, and $\mathscr{A}_a(x)$ is the local generator of axial transformations. We have defined the vector and axial gauge fields $(\mcV,\mcA)$ by $\cA_{\textrm{\tiny L}} \equiv \mcV - \mcA$ and $\cA_{\textrm{\tiny R}} \equiv \mcV + \mcA$. The anomaly also leads to the (non)-conservation law $D_\mu J_a^\mu(x)_{\textrm{cons}} = G_a[\mcV,\mcA]$. As a consequence, the chiral anomaly has effects in the hydrodynamic constitutive relations, as it has been already discussed. The Bardeen form of the non-Abelian anomaly is~\cite{Bardeen:1969md}
\begin{eqnarray}
\hspace{-1.2cm}&&G_a[\mathcal{V}, \mathcal{A}] = \frac{i N_c}{16 \pi^2} \epsilon^{\mu \nu  \rho \sigma} \times  \nonumber \\
\hspace{-1.2cm}&&\times \text{Tr} \Bigl\{
t_a \bigl[ \mcV_{\mu \nu} \mcV_{\rho \sigma} + \tfrac{1}{3} \mcA_{\mu \nu} \mcA_{\rho \sigma} - \tfrac{32}{3} \mcA_\mu \mcA_\nu \mcA_\rho \mcA_\sigma  \bigl] \nonumber \\ 
\hspace{-1.2cm}&&\quad + \tfrac{8}{3} i  (\mcA_\mu \mcA_\nu \mcV_{\rho \sigma}  + 
\mcA_\mu \mcV_{\rho \sigma} \mcA_\nu  + \mcV_{\rho \sigma} \mcA_\mu \mcA_\nu) \bigr] \Bigr\} \,,
\end{eqnarray}
%
where $N_c$ is the number of colors, while $(\mcV_{\mu \nu}, \mcA_{\mu \nu})$ are the field strengths. $G_a$ includes triangle, square and pentagon one-loop diagrams, in contrast to the Abelian case in which only triangle diagrams~contribute.

\subsection{Constitutive relations}
\label{subsec:Non_Sbelian_Const}

The solution of the anomaly equation can be found by using differential geometry methods based on the Chern-Simons effective action, with the result~\cite{Manes:2018llx,Manes:2019fyw}
\begin{eqnarray}
\hspace{-0.9cm}&&\Gamma[V, A, G] = -\frac{N_c}{32 \pi^2}  \int dt \,  d^3 x  \sqrt{g_3} \, 
 \epsilon^{i j k} \times  \nonumber \\  
\hspace{-0.9cm}&& \quad \times \; \text{Tr} \biggl\{  \frac{32}{3} i \, V_0 A_i A_j A_k  + \frac{4}{3} (A_0 A_i + A_i A_0) A_{j k}    \nonumber \\ 
\hspace{-0.9cm}&& \quad + 4  (V_0 A_i + A_i V_0) V_{j k}   + \frac{8}{3} \bigl(A_0^2 + 3 V_0^2 \bigr) A_i \partial_j a_k \biggr\}  \,.
\end{eqnarray}
We have neglected the terms related to the mixed gauge-gravi\-tational anomaly $\sim C_2$, as these contributions demand a careful study of the Riemann tensor effects in the anomaly polynomial, see e.g. Ref.~\cite{Nair:2011mk}.

In the $(uds)$ flavor sector of QCD, the conserved charges are the baryon number $B$, electric charge $Q$, and strangeness $S$. Then, instead of working in the basis of the generators of the Cartan subalgebra for $N_f = 3$, $\{t_0, t_3, t_8\}$, it is more convenient to work in the $\{B,Q,S\}$ basis, for which we can take the following background $V_\mu(\vec{x}) = V_{B \,  \mu}(\vec{x}) B + \, V_{Q \, \mu}(\vec{x}) Q + \, V_{S\, \mu}(\vec{x}) S$, $A_0   = A_{B} \, B$ and $A_i = 0$. Then we can distinguish between the three vector currents  $J^\mu_{B} = \overline\Psi \gamma^\mu B \Psi$, $J^\mu_{\eem} = e \overline\Psi \gamma^\mu Q \Psi$ and $J^\mu_{S} = \overline\Psi \gamma^\mu S \Psi$, corresponding to the baryonic, electromagnetic and strangeness currents, respectively. In addition, we can define the corresponding chemical potentials as $\mu_q = e^{-\sigma} {\cal V}_{q \, 0}\; (q = B,Q,S)$ and $\mu_5 = e^{-\sigma} {\cal A}_{0\, 0}$. Here, $\mu_5$ controls the chiral imbalance of the system~\cite{Gatto:2011wc}.

The covariant currents are defined by adding to the consistent currents the Bardeen-Zumino (BZ) terms, i.e. $J^\mu_\cov = J^\mu_\cons + J^\mu_{\BZ}$~\cite{Bardeen:1984pm}. These are the physically relevant currents, as can be argued using the notion of anomaly inflow~\cite{Callan:1984sa}. To compute the constitutive relations, let us assume that the electromagnetic field is the only propagating field. Then, we can define the physical magnetic field as ${\mathcal B}^\mu = \frac{1}{2} \epsilon^{\mu\nu\alpha\beta} u_\nu  \mbhV_{\alpha\beta}$, where the physical potential is $\mbhV_\mu$ and its KK invariant form is $\mbbV_\mu$, i.e. $\mbbV_0 = \mbhV_0$ and $\mbbV_i = \mbhV_i - a_i \mbhV_0$. Then one has $V_{B\, \mu} = 0 = V_{S\, \mu}$ and $V_{Q\, \mu} = e \mbbV_\mu$, and the constitutive relations write~\cite{Manes:2022zrl,Megias:prep}
\begin{align}
\langle  J^\mu_{\eem} \rangle_{\cov} &= \frac{e^2 N_c}{3 \sqrt{6} \pi^2} \mu_5 {\mathcal B}^\mu  \,, \\
q^\mu &= \frac{N_c}{3\sqrt{6} \pi^2} \mu_5 \left[ e \mu_Q {\mathcal B}^\mu + \left( \mu_Q^2 - \frac{1}{4} \mu_5^2 \right) \omega^\mu \right] \,,
\end{align}
where $\langle T^{\mu\nu}\rangle = u^\mu q^\nu + u^\nu q^\mu$. Notice that $\langle  J^\mu_{\eem} \rangle_{\cov}$ receives contribution only from the chiral magnetic conductivity. The absence of a chiral vortical effect in the $\U(3)_V \times \U(3)_A$ case contrasts with the situation in the Abelian $\U(1)_V \times \U(1)_A$ model, cf. Sec.~\ref{subsec:first_order} and Refs.~\cite{Landsteiner:2012kd,Jensen:2013vta}.

\section{Conclusions}
\label{sec:Conclusions}

We have studied the anomaly-induced transport effects in relativistic fluids by using the EPF formalism. By construction, this method can only account for non-dissipative effects, i.e. transport coefficients multiplying quantities that survive in equilibrium. In particular, we have characterized the effects induced by external magnetic fields and fluid vorticity. In the Abelian case, the non-dissipative contributions at first order are  ${\cal P}$-odd and  ${\cal T}$-even. However, the situation is slightly different at second order, where the ${\cal P}$-odd coefficients vanish, and the nonzero non-dissipative coefficients turn out to be ${\cal P}$-even and ${\cal T}$-even. In the case of non-Abelian anomalies, we have found that there are contributions to the constitutive relations at first order from the physical magnetic field, but no contribution from the vorticity. While the present study is relevant for the chiral symmetric phase of QCD at high temperatures, the computation has been extended in Refs.~\cite{Manes:2018llx,Manes:2019fyw} to the case of spontaneous symmetry breaking, leading to relevant information about the hydrodynamics of the Goldstone bosons interacting with external fields, with application to QCD at low temperatures. Finally, let us remark that this formalism can be used in a wide variety of systems, including other sectors of the Standard Model~\cite{Brauner:2012gu}, superfluids~\cite{Hoyos:2014nua}, and condensed matter systems~\cite{Basar:2013iaa,Landsteiner:2013sja}.

\section*{Acknowledgements} 
This work is based on Ref.~\cite{Megias:2014mba}, co-authored with
M. Valle, and Refs.~\cite{Manes:2018llx,Manes:2019fyw}, co-authored
with J.L. Ma\~nes, M. Valle and M.\'A. V\'azquez-Mozo. I would like to
thank them for collaboration and enlightening discussions. I also
thank the ICTP South American Institute for Fundamental Research
(SAIFR), S\~ao Paulo, Brazil, and its Program on New Directions in
Particle Physics 05-23/09/2022, for hospitality and partial financial
support during the process of writing this manuscript. This work is
supported by the project PID2020-114767GB-I00 financed by
MCIN/AEI/10.13039/501100011033, by the FEDER/Junta de
Andaluc\'{\i}a-Consejer\'{\i}a~de~Econom\'{\i}a~y Conocimiento
2014-2020 Operational Programme under Grant A-FQM-178-UGR18, by Junta
de Andaluc\'{\i}a under Grant FQM-225, and by the Consejer\'{\i}a~de
Cono\-cimiento, Investigaci\'on y Universidad of the Junta de
Andaluc\'{\i}a and European Regional Development Fund (ERDF) under
Grant SOMM17/6105/UGR. This research is also supported by the Ram\'on
y Cajal Program of the Spanish MCIN under Grant RYC-2016-20678.


\begin{thebibliography}{99}
\expandafter\ifx\csname url\endcsname\relax
  \def\url#1{\texttt{#1}}\fi
\expandafter\ifx\csname urlprefix\endcsname\relax\def\urlprefix{URL }\fi
\expandafter\ifx\csname href\endcsname\relax
  \def\href#1#2{#2} \def\path#1{#1}\fi

\bibitem{Fukushima:2008xe}
K.~Fukushima, D.~E. Kharzeev, H.~J. Warringa,
\newblock  Phys. Rev. {\bf D78} (2008) 074033.

\bibitem{Son:2009tf}
D.~T. Son, P.~Surowka, 
\newblock Phys. Rev. Lett. {\bf 103} (2009) 191601.

\bibitem{Stephanov:2012ki}
M.~A. Stephanov, Y.~Yin, 
\newblock Phys. Rev. Lett. {\bf 109} (2012) 162001.

\bibitem{Landsteiner:2012kd}
K.~Landsteiner, E.~Meg\'{\i}as, F.~Pena-Benitez, 
\newblock Lect. Notes Phys. {\bf 871} (2013) 433--468.

\bibitem{Erdmenger:2008rm}
J.~Erdmenger, M.~Haack, M.~Kaminski, A.~Yarom, 
\newblock JHEP {\bf 01} (2009) 055.

\bibitem{Banerjee:2012iz}
N.~Banerjee, J.~Bhattacharya, S.~Bhattacharyya, S.~Jain, S.~Minwalla,
  T.~Sharma, 
\newblock JHEP {\bf 09} (2012) 046.

\bibitem{Jensen:2013kka}
K.~Jensen, R.~Loganayagam, A.~Yarom,
\newblock JHEP {\bf 05} (2014) 134.

\bibitem{Bhattacharyya:2013lha}
S.~Bhattacharyya, 
\newblock JHEP {\bf 08} (2014) 165.

\bibitem{Megias:2014mba}
E.~Meg\'{\i}as, M.~Valle, 
\newblock JHEP {\bf 11} (2014) 005.

\bibitem{Landsteiner:2011cp}
K.~Landsteiner, E.~Meg\'{\i}as, F.~Pena-Benitez,
\newblock Phys. Rev. Lett. {\bf 107} (2011) 021601.

\bibitem{Landsteiner:2011iq}
K.~Landsteiner, E.~Meg\'{\i}as, L.~Melgar, F.~Pena-Benitez, 
\newblock JHEP {\bf 09} (2011) 121.

\bibitem{Banerjee:2008th}
N.~Banerjee, J.~Bhattacharya, S.~Bhattacharyya, S.~Dutta, R.~Loganayagam,
  P.~Surowka, 
\newblock JHEP {\bf 01} (2011) 094.

\bibitem{Giannotti:2008cv}
M.~Giannotti, E.~Mottola, 
\newblock Phys. Rev. {\bf D79} (2009) 045014.

\bibitem{Bhattacharyya:2013ida}
S.~Bhattacharyya, J.~R. David, S.~Thakur, 
\newblock JHEP {\bf 01} (2014) 010.

\bibitem{Moore:2012tc}
G.~D. Moore, K.~A. Sohrabi, 
\newblock JHEP {\bf 11} (2012) 148.

\bibitem{Megias:2013joa}
E.~Meg\'{\i}as, F.~Pena-Benitez, 
\newblock JHEP {\bf 05} (2013) 115.

\bibitem{Bardeen:1969md}
W.~A. Bardeen, 
\newblock Phys. Rev. {\bf 184} (1969) 1848--1857.

\bibitem{Manes:2018llx}
J.~L. Ma\~nes, E.~Meg\'{\i}as, M.~Valle, M.~\'A. V\'azquez-Mozo,
\newblock JHEP {\bf 11} (2018) 076.

\bibitem{Manes:2019fyw}
J.~L. Ma\~nes, E.~Meg\'{\i}as, M.~Valle, M.~\'A. V\'azquez-Mozo, 
\newblock JHEP {\bf 12} (2019) 018.

\bibitem{Nair:2011mk}
V.~P. Nair, R.~Ray, S.~Roy, 
\newblock Phys. Rev. {\bf D86} (2012) 025012.

\bibitem{Gatto:2011wc}
R.~Gatto, M.~Ruggieri,
\newblock Phys. Rev. {\bf D85} (2012) 054013.

\bibitem{Bardeen:1984pm}
W.~A. Bardeen, B.~Zumino, 
\newblock Nucl. Phys. {\bf B244} (1984) 421--453.

\bibitem{Callan:1984sa}
C.~G. Callan, Jr., J.~A. Harvey, 
\newblock Nucl. Phys. {\bf B250} (1985) 427--436.

\bibitem{Manes:2022zrl}
J.~L. Ma\~nes, E.~Meg\'\i{}as, M.~Valle, M.~A. V\'azquez-Mozo, 
\newblock EPJ Web Conf. {\bf 258} (2022) 10006.

\bibitem{Megias:prep}
E.~Meg\'{\i}as, M.~\'A. V\'azquez-Mozo, 
\newblock to appear (2022).

\bibitem{Jensen:2013vta}
K.~Jensen, P.~Kovtun, A.~Ritz, 
\newblock JHEP {\bf 10} (2013) 186.

\bibitem{Brauner:2012gu}
T.~Brauner, O.~Taanila, A.~Tranberg, A.~Vuorinen,
\newblock JHEP {\bf 11} (2012) 076.

\bibitem{Hoyos:2014nua}
C.~Hoyos, B.~S. Kim, Y.~Oz, 
\newblock JHEP {\bf 10} (2014) 127.

\bibitem{Basar:2013iaa}
G.~Basar, D.~E. Kharzeev, H.-U. Yee,
\newblock Phys. Rev. {\bf B89}~(3) (2014) 035142.

\bibitem{Landsteiner:2013sja}
K.~Landsteiner,
\newblock Phys. Rev. {\bf B89}~(7) (2014) 075124.

\end{thebibliography}

\end{document}